\documentclass[aps,pra,twocolumn,groupedaddress,showpacs,10pt,longbibliography]{revtex4-1}

\usepackage{graphicx}
\usepackage{dcolumn}
\usepackage{bm}
\usepackage{setspace}
\usepackage{color}
\usepackage{overpic}
\usepackage{sidecap}
\usepackage{braket}

\usepackage[english]{babel}
\draft

\begin{document}

\title{Vortex reconnections and rebounds in trapped atomic Bose--Einstein condensates}

\author{Simone Serafini$^1$}
\author{Luca Galantucci$^2$}
\author{Elena Iseni$^1$}
\author{Tom Bienaim\'e$^1$}
\author{Russell N. Bisset$^1$}
\author{Carlo F. Barenghi$^2$}
\author{Franco Dalfovo$^1$}
\author{Giacomo Lamporesi$^1$}
\email[]{giacomo.lamporesi@ino.it}
\author{Gabriele Ferrari$^1$}

\affiliation{$^1$ INO--CNR BEC Center and Dipartimento di Fisica, Universit\`a di Trento, 38123 Povo, Italy \\
$^2$ Joint Quantum Centre (JQC) Durham--Newcastle, and School of Mathematics and Statistics, 
Newcastle University, Newcastle upon Tyne, NE1 7RU, United Kingdom}

\begin{abstract}
Reconnections and interactions of filamentary coherent structures 
play a fundamental role in the dynamics of fluids,
redistributing energy and helicity among the length scales 
 and inducing fine-scale turbulent mixing. 
Unlike ordinary fluids, where vorticity is a continuous field, 
in quantum fluids vorticity is concentrated into discrete (quantized) 
vortex lines turning vortex reconnections into isolated events, 
making it conceptually  easier to study. Here we report experimental and numerical 
observations of three-dimensional quantum vortex interactions in a 
cigar-shaped atomic Bose--Einstein Condensate. In addition to 
standard reconnections,  already numerically 
and experimentally observed in homogeneous systems 
away from boundaries, we show that double reconnections, rebounds
and ejections can also occur as a consequence of the non-homogeneous, 
confined nature of the system.
\end{abstract}

\maketitle

\section{\label{intro}Introduction}
The interaction and reconnection of filaments are key aspects in the description of the 
dynamics of fluids \cite{Kida94,Barenghi01,Schwarz88}, plasmas 
\cite{Priest07,Che11,Cirtain13}, nematic liquid crystals \cite{Chuang91},
macromolecules \cite{Sumners95} (including DNA \cite{Vazquez04}) and 
optical beams \cite{Dennis10,Berry12}.
In quantum fluids, vortices are topological defects of the 
system's order parameter, around which the circulation of the 
velocity field is quantized \cite{Onsager49,Feynman55,Vinen61,Donnelly91}. 
Their discrete filamentary nature makes quantum fluids an ideal 
setting for the study of vortex interactions and reconnections. In particular, 
reconnections trigger a turbulent energy cascade \cite{Barenghi14} in which
vortex lines self-organize in bundles \cite{Baggaley12} creating the 
same Kolmogorov distribution of kinetic energy over the length 
scales, signature of a cascade mechanism which is observed in ordinary turbulence 
\cite{Nore97,Maurer98,Skrbek12,Barenghi14}. 
Cascade processes are central in turbulent motions.
A related cascade of wave-like excitations
was in fact recently observed in the momentum distribution \cite{Navon16},
with an exponent consistent with predictions of wave-turbulence theory \cite{zakharov2012}.
Reconnection events also impact 
on the evolution of the flow's topology \cite{Kleckner16}, redistributing 
helicity among length scales \cite{Clark16,Scheeler14}. Finally, in the low-temperature 
limit, reconnections are the ultimate process of dissipation of superfluid 
kinetic energy since they trigger a Kelvin wave cascade \cite{Kozik04,Kozik06}
that turns incompressible kinetic energy into acoustic modes \cite{Leadbeater01}, 
hence heating. Previous experimental \cite{Bewley08,Fonda14}, 
theoretical \cite{Nazarenko03}
and numerical \cite{Koplik93,Dewaele94,Tebbs11,Zuccher12,Kerr11,Villois16b,Rorai16} 
studies of reconnections
have been performed in homogeneous systems away from boundaries. 

Here we focus on elongated Bose--Einstein condensates (BECs) of ultracold atoms confined by 
magnetic harmonic potentials, ideal systems which allow for different
regimes of three-dimensional (3D) vortex-vortex interactions in the
close presence of boundaries. 
Anisotropic boundaries induce vortical filaments to preferentially 
align along the shortest direction, minimising energy. 
In flat, cylindrically symmetric, disk-shaped condensates, 
vortices are the shortest when aligned along 
the axis of symmetry, moving along two-dimensional trajectories
clockwise or anti-clockwise, depending on their 
sign \cite{Anderson00,Weiler08,Freilich10,Neely10,Torres11,Navarro13}. Instead, vortices in 
cylindrically symmetric, cigar-shaped condensates are the 
shortest when they lie on radial planes. Moreover, the boundaries
affect the structure of the vortical flow  \cite{Brand02,Komineas03,Donadello14,Tylutki15} in 
such a way that two  vortices only interact when their minimum
distance is within a range of the order of the transverse size of the condensate.  

\begin{figure*}[t]
\includegraphics[width=1\textwidth]{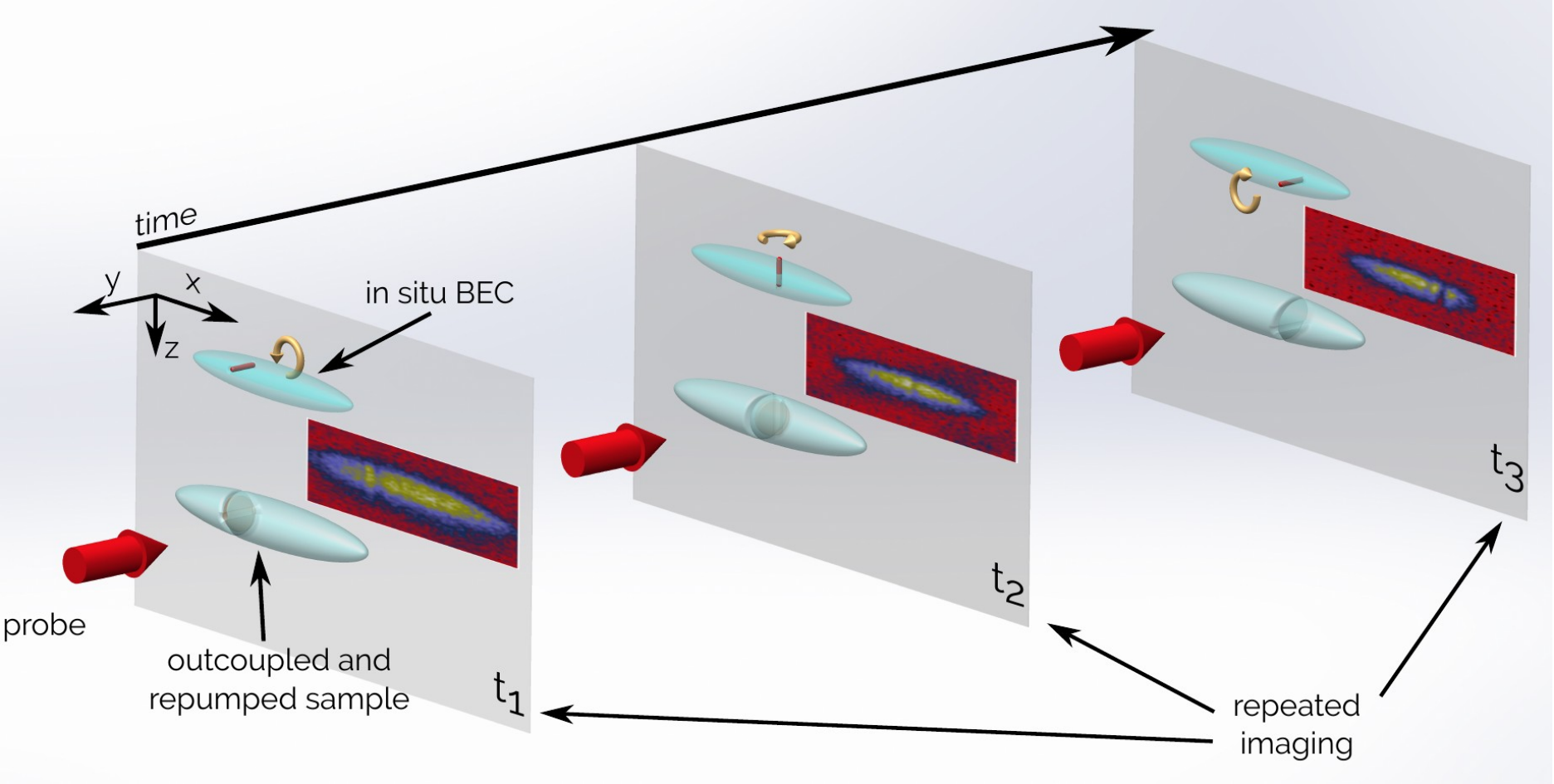}
\caption{
Sketch of an imaging sequence. 
A trapped condensate (smaller light-blue ellipsoid) contains a transverse vortex line that moves 
and rotates around the trap center; the direction of the atomic flow around the vortex 
filament is indicated by the yellow arrow. A small fraction of atoms is repeatedly extracted,
typically every 12 ms;  these atoms expand and fall in the gravity field, and are 
imaged in absorption by a probe laser beam after they are spatially separated from the 
trapped condensate. Each absorption image contains the essential features associated 
with the vortex lines.}  
\label{fig:1}
\end{figure*}

In the present work, an innovative imaging technique, exploiting self-interference effects of 
outcoupled atoms, is introduced in order to extract both the position and orientation
of 3D vortex lines from a temporal sequence of absorption images. We then combine experiments 
and numerical Gross-Pitaevskii (GP) simulations to study the interaction between two vortex lines approaching at 
various relative speeds and angles. Our experiments and simulations show that the interaction
between vortex lines in a finite system is rather different from the one in infinite uniform 
superfluids. Boundary-induced effects, such as rebounds, double reconnections, and ejections, 
are here discussed in details. These types of processes may play an important role in the dynamics
of trapped condensates in multi-vortex and turbulent-like configurations, and, on a wider 
perspective, they  can represent novel keys for better understanding the behavior of superfluids
near boundaries. 

\section{\label{experiment}Experiment}
\subsection{Preparation of BECs with vortices}
Experimentally, we confine sodium atoms in an elongated cigar-shaped harmonic magnetic trap 
with axial and radial frequencies $\omega_x/2\pi=9.2$~Hz and  $\omega_{\perp}/2\pi=92$~Hz, 
respectively. By means of a radio-frequency forced evaporation the cold atomic sample undergoes 
the BEC transition and, in the end, condensates containing about $N_0=2\times 10^7$ atoms 
and a negligible thermal fraction ($T<150$~nK, $T_c \simeq 500$~nK) are obtained. Thanks to the 
Kibble--Zurek mechanism \cite{Kibble76,Zurek96} the temperature quench through the BEC transition \cite{Weiler08,Freilich10,LamporesiKZM13,Donadello16} produces different phase domains in the order parameter 
of the system that quickly evolve into topological defects. In our geometry, these defects are vortex lines mainly 
oriented in the transverse direction, as those predicted in \cite{Brand02,Komineas03} and characterized in 
 \cite{Donadello14}. Similar vortices can be obtained as decay products of  phase imprinted dark solitons in 
a BEC \cite{Becker13} or a Fermi superfluid gas \cite{Ku14,Ku16}. Here we use a cooling rate of $4\ \mu$K/s 
in order to produce, on average, two vortices in each condensate at the time when the observation starts, 
about $250$ ms after the phase transition. 
Such vortices move in the non-rotating condensate and can be directly imaged in 
real-time \cite{Freilich10,Ramanathan14,Serafini15}. 
In comparison, individual vortex visualisation in superfluid helium is more intrusive, 
requiring tracer particles whose diameter is about $10^4$ times larger than the 
vortex core \cite{Bewley06}.\\

\begin{figure}[t!]
\includegraphics[width=1\columnwidth]{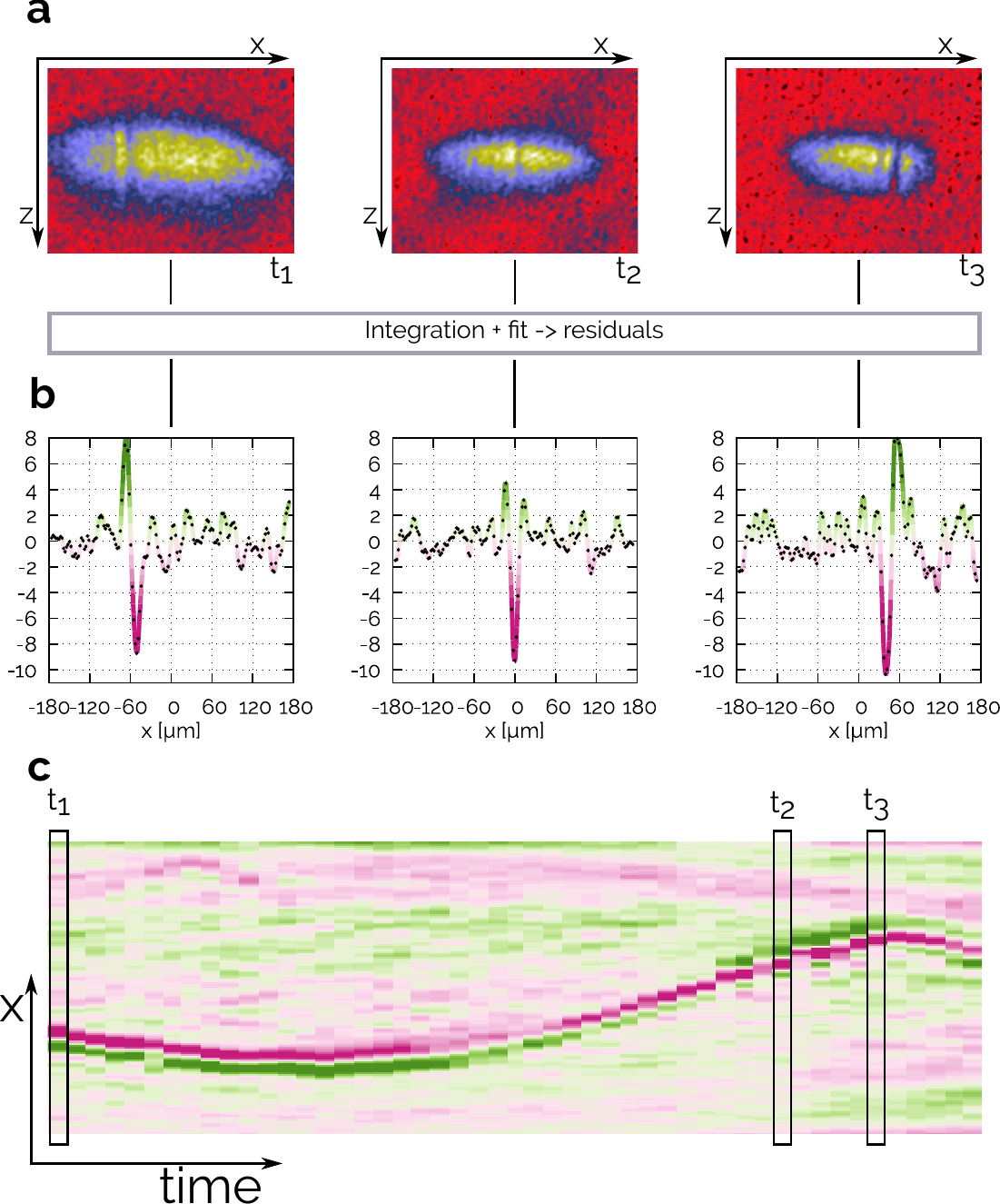}
\caption{
(a) Examples of absorption images of the outcoupled atoms (the same as in Fig.~\ref{fig:1}). 
The vortex axial position is clearly visible.  (b) After integrating radially and fitting the 
absorption images, we determine the residuals, which exhibit minima (pink) and maxima (green)  
due to interference effects among atoms that are outcoupled from the trapped condensate at different places 
and times.  (c) Full temporal sequence of residuals for a given condensate, 
showing the real-time evolution of a vortex which moves axially and rotates around the $x$ axis, from 
an initial orientation along $y$ (green-pink)  at $t_1$ to an orientation along $z$ (green-pink-green) 
in $t_2$ and then along $-y$ (pink-green) at $t_3$. The relation between the shape of the residuals
and the orientation of the vortex is extracted from numerical simulations.} 
\label{fig:2}
\end{figure}

\subsection{Sample extraction and real-time imaging}

A new imaging method allows us to follow the vortex dynamics in real-time, as
sketched in Fig.~\ref{fig:1}.
Similar to \cite{Freilich10,Ramanathan14}, a small sample of the atomic system ($\sim 10^5$ atoms) is repeatedly extracted from the BEC every 12 ms (up to 75 times). The outcoupled atoms freely expand and fall under the effect of gravity.
Each partial extraction is implemented by coupling the trapped state $| F=1, m_F=-1 \rangle$ to the 
non-magnetic one $| 1, 0 \rangle$ with a radio frequency (\textit{rf}) field. 
The energy difference between the two states is spatially dependent because of the inhomogeneity of the trapping potential (see Appendix A).

The novelty of our technique is represented by the fact that the {\it rf} field is frequency-swept linearly in time in order to match the resonant condition at different positions throughout the BEC, from top to bottom. 
An important point to note is that the phase of the released atoms evolves more slowly because they do not feel the trapping potential. As a consequence, the wave function of the outcoupled atoms experiences constructive or destructive self-interference effects, depending on the phase difference accumulated between the early-released (upper) and late-released (lower) atoms, and how this relates to the {\it in situ} phase on different sides of the vortex core.
We use the GP equation to simulate the radio frequency extraction in order 
to determine how a vortex with given position and orientation in 
the trapped BEC manifests itself 
in the observed density distribution of the
outcoupled atoms after expansion (see Appendix C).

A microwave field remains on to transfer the extracted atoms from $| 1, 0 \rangle$ to $|2,0\rangle$, 
which is detectable with the probe light. The resonant condition for the transfer is matched at $z_r\approx 280\,\mu$m below the trapped BEC, far enough to leave it unaffected.
We probe the extracted atoms via standard absorption imaging after $13$~ms of
total time of flight at $z_i\approx 830\,\mu$m below the trap center.
Such a time of flight is enough for vortices to become visible with our imaging resolution.

\begin{figure}[t!]
\centering
\includegraphics[width=0.82\columnwidth]{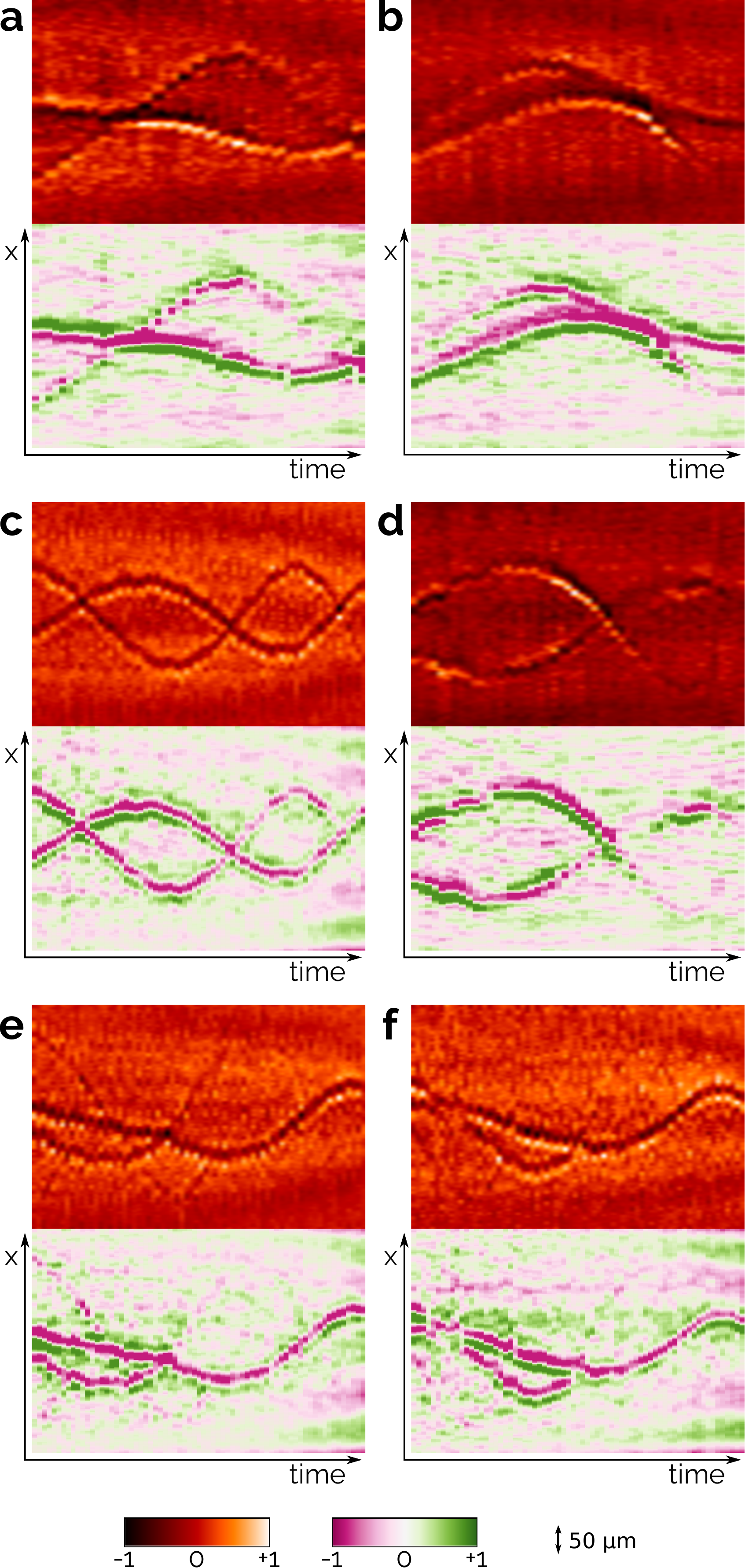}
\caption{Examples of different interaction mechanisms observed in 
the case of two approaching vortices. Each temporal sequence is shown twice with two different color palettes; the red palette enhances the contrast, so that also vortices close to the edges can be seen, whereas the pink-green palette better illustrates the vortex orientation in the radial plane. (a,b) Vortices approach and bounce 
back; (c) their axial trajectories intersect preserving visibility and orientation; (d) they cross producing 
sudden changes of visibility; (e,f) the visibility of 
one vortex is almost completely lost after interacting with the other.}
\label{fig:3}
\end{figure}

\subsection{Data analysis}
Each absorption image (Fig. \ref{fig:2}a) is integrated radially along the $z$ axis and the axial profile is obtained. By fitting the latter, we calculate the density residuals (see Fig. \ref{fig:2}b). This procedure is performed on each extraction and then the full temporal sequence is reconstructed in order to follow the vortex trajectories in the trapped condensate, as in Fig. \ref{fig:2}c. 

Thanks to the above mentioned self-interference effect, if a vortex is present, the density residuals show a strong local deviation from the unperturbed distribution, as in Fig.~\ref{fig:2}b, and the fit allows us to extract information on the vortex axial position, as well as on its orientation in the radial plane at any given time (see Appendix C for details). 

Figure \ref{fig:3} shows examples of the temporal evolution of 
the density-residual profiles in BECs containing two vortices. Two different color palettes are used in order to extract different pieces of information.
The red palette best highlights the trajectory contrast. One can track the vortex axial location in time and hence determine the orbit amplitude and the axial velocity. Notice that in some cases, very faint trajectories (corresponding to vortices close to the BEC surface) can also be seen.
It is also possible to understand how the vortex line is oriented in the radial plane and how it rotates about the long axis of the condensate. The diverging pink-green palette helps to visualize the shape of the density modulation from which one can better track the vortex orientation in time. From numerical simulations we infer that, at least when the orbiting parameter is not too large, the vorticity points along $y$ if the interference pattern is green-pink  along $x$ (see row (b) in Fig.~\ref{fig:2}); its anti-vortex configuration, oriented toward $-y$, corresponds to a pink-green pattern; the symmetric pattern green-pink-green is obtained when the vortex is aligned perpendicularly to the imaging direction, a vortex oriented along $+z$ providing the same density residual as one oriented along $-z$.

\section{\label{simulations}Numerical simulations}
In order to gain closer insight into vortex interactions, we perform 
numerical simulations by using the Gross--Pitaevskii equation \cite{Pitaevskii16,Dalfovo99}
for a BEC at $T=0$. Temperature effects are expected to be small. In a previous work \cite{Serafini15},  
we have already observed that the dynamics of single vortices is very weakly 
affected by thermal excitations. 
This is expected to be true also for vortex-vortex interaction 
processes occurring in the central region of our BEC, 
where the thermal density is negligible.
There is also evidence that thermal excitations do not affect the
rapid motion of vortex lines during the reconnections \cite{Allen14}.

We track the vortices by employing an algorithm based on the 
pseudo-vorticity vector, achieving sub-grid resolution (see Appendix B).   
Since the experimental BECs are too large for our computational
resources, we simulate smaller BECs ($\sim 4\times 10^5$ atoms); 
this implies a reduction of the ratio $R_\perp/\xi$ 
by a factor of three, where $\xi$ is the 
healing length and $R_\perp$ is the transverse Thomas--Fermi radius. 
However, such a difference does not affect the qualitative comparison between experiment and simulations.  

If we imprint a single straight vortex line off-center on 
a radial plane, we find that it
orbits around the center of the condensate \cite{Anderson00,Freilich10}
along an elliptical orbit which is orthogonal to the vortex line itself.
The orbit, which is a trajectory of constant energy \cite{Svidzinsky00}
and an isoline of the trapping potential, is
uniquely determined by the orbit parameter 
$\chi=r_0/R_\perp=x_0/R_x$, where $r_0$ and $x_0$ correspond to 
the radial and axial semi-axes of the ellipse, while $R_x$ 
is the axial Thomas--Fermi radius. 
The orbital period is maximum when the vortex moves on a very small orbit ($\chi\ll 1$) 
and corresponds to $T_0=8\pi \mu/[3\hbar\omega_\perp \omega_x \ln(R_\perp/\xi)]$ \cite{Serafini15}, while it
decreases with increasing $\chi$ \cite{Svidzinsky00,Lundh00,Sheehy04,Fetter09}.

If instead we imprint two transverse vortices in a given BEC, we find
that the evolution can be 
divided into two stages. In the first stage, when the axial separation
of the vortices is larger than $R_{\perp}$,
the vortices move almost independently; 
in the second stage, when the axial separation becomes smaller than
$R_{\perp}$, we observe a significant interaction
which seems to be determined mainly
by the relative 
orientation $\theta_{\mathrm{rel}}$ and velocity  $v_{\mathrm{rel}}$
when they start interacting.

\begin{figure*}
\centering
\includegraphics[width=0.65\textwidth]{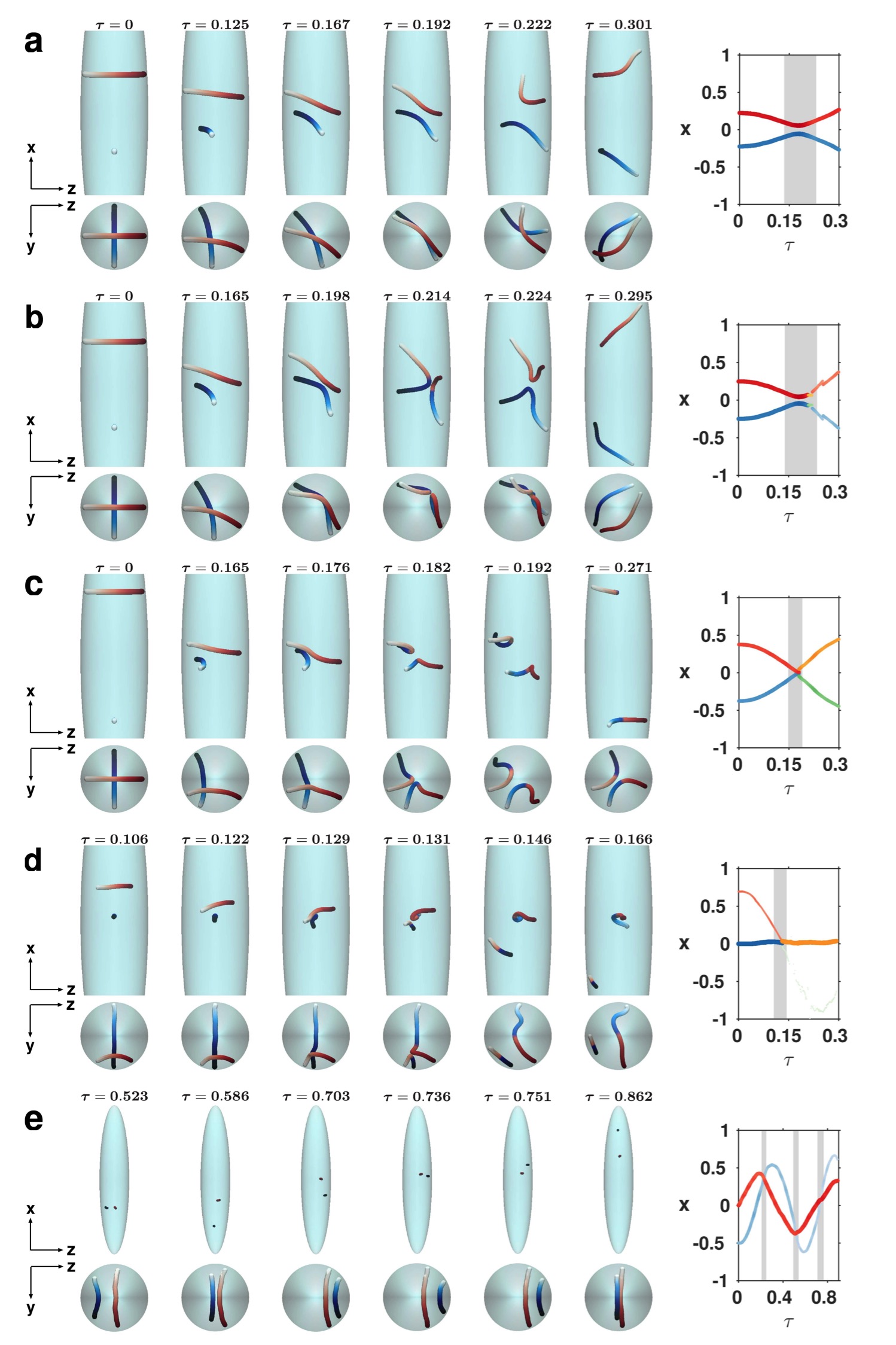}
\vspace{-0.2cm}
\caption{The first six columns show radial and axial snapshots 
from the GP simulations of two interacting vortex lines. 
On the right, the axial coordinate $x$ (in units of $R_x$) of the center of vorticity of each vortex
is plotted {\it vs.} normalized time $\tau=t/T_0$. 
Initial line colors (red/blue) help identify vortices in the snapshots until they reconnect.
After the first reconnection, line colors switch to orange/green and again to red/blue if a second reconnection occurs.
Line transparency indicates how visible vortices are expected to be, given their orbit amplitude 
(see Appendix B for further details on line transparency). 
The grey region highlights the interaction interval 
where the minimum distance between the vortices
is smaller than $R_\perp$.
(a-c) Perpendicular vortices are imprinted on opposite radial planes
with corresponding orbit parameters $\chi=0.22,\, 0.25, \, 0.375$, respectively:
(a) illustrates a vortex rebound; 
(b) shows the double reconnection interaction, with reconnections 
occurring at $\tau=0.208$ and $\tau=0.221$ (see Appendix B for a zoom on the double reconnection event); 
(c) depicts a single reconnection occurring at $\tau=0.179$,
with the consequent triggering of Kelvin waves.
(d) illustrates a non-symmetrical reconnection (at $\tau=0.130$)
between a vortex imprinted on the central plane of the condensate
through its center (blue) and a vortex (red) 
imprinted orthogonally to the first one with a large orbit parameter $\chi=0.7$.
One of the reconnected vortices lies on an even wider orbit (larger $\chi$), 
where the BEC density is lower and its visibility 
becomes consequently greatly reduced.
(e) describes the orbiting dynamics between two parallel vortices 
imprinted on different orbits ($\chi=0.33,\, 0.5$). 
Notice that in (a-c) the first snapshot corresponds to $\tau=0$, 
whereas in (d,f) the snapshots are all later in time.  }
\label{fig:4}
\end{figure*}

We first perform simulations in which two orthogonal vortices are initially 
imprinted in radial planes at opposite axial positions $\pm x_0$, 
see Figs.~\ref{fig:4}a-c (orthogonality is 
chosen because of its maximal dissimilarity with respect to flat 2D systems). 
Different $x_0$ values are chosen, corresponding to different orbit parameters 
$\chi$ and hence to different impact velocities. 
The early stage can be described as the combination of two single-vortex 
motions on mutually perpendicular elliptical orbits. 

In fact, in an elongated condensate, the superfluid flow of each vortex 
becomes negligible at distances of the order of $R_{\perp}$ from the line, as 
can be verified by solving the stationary GP equation. This means that when two 
vortices are at distances larger than $\sim R_{\perp}$, they behave 
as non-interacting objects, as indeed observed in time-dependent GP simulations. 
This is crucial in order to interpret and classify the vortex-vortex 
interaction as a collision with well-defined initial and final 
velocities and orientations. In a different 3D geometry, it would be very difficult
to define and control a global ``relative velocity and orientation" of a vortex line. 
If a non-rotating condensate is confined in a spherical potential, or is uniform, for instance, the 
distance between two vortices and their relative velocity and orientation 
could be defined only locally: vortices do not possess a preferred 
orientation, they can be easily bent, and each piece of vortex is affected by a
long-range interaction with all other vortices in the condensate.
Our geometry instead naturally provides  well-defined collision events, such as 
reconnecting or bouncing lines, occurring in a narrow interaction region. 

Only when the minimum distance between the vortices becomes of the order 
of $R_{\perp}$, the vortices start rotating in the radial plane, attempting to arrange 
themselves in the preferred (energy-conserving) anti-parallel configuration, as shown in Figs.~\ref{fig:4}a-c. 
The axial motion of the vortices towards each other, 
driven by the inhomogeneous density, is faster
if the vortices are close to the condensate's boundary
\cite{Svidzinsky00,Sheehy04,Fetter09}. 
The anti-parallel configuration which the vortices attempt to achieve induces them to drift
radially towards the radial center of the condensate. This drift is similar to the 
well-known self-induced motion of a pair of straight anti-parallel vortex lines \cite{Neely10,Torres11,Middelkamp11}
in a homogeneous condensate. The balance between the radial and axial motions 
which we have described determines the features of the second stage of the interaction.

Briefly, if the axial collision velocity is sufficiently high (\textit{i.e.}, if the vortex 
lines start interacting in a region sufficiently close to the boundary \cite{Svidzinsky00,Sheehy04,Fetter09}) 
the two vortices tend to reconnect before 
reaching the center of the condensate, as in Fig.~\ref{fig:4}c. 

Vice versa, if the interaction begins in a region sufficiently close to the $x$-axis, 
the radial motion of the vortex lines is fast enough (with respect to 
the axial motion) to get past the radial
center of the condensate where they move axially away from each other due to the reversed velocity 
field induced by the inhomogeneous density: a rebound takes place, 
as in Fig.~\ref{fig:4}a. 

An intermediate regime occurs if, while drifting radially away from the 
boundary of the condensate towards the center, the minimum distance between the vortices
in the central region of the condensate is sufficiently small: in this case
a double reconnection \cite{Berry12} occurs. This happens for instance
in the sequence in Fig.~\ref{fig:4}b, where the two vortex lines touch at a point 
and exchange their tails both at $\tau=0.208$ and $\tau=0.221$, expressed in units of the precession period $T_0$
(see Fig.~\ref{Fig:Double_Recon} in Appendix B for a more detailed illustration of the double reconnection dynamics).

In addition to the simulations with chemical potential $\mu=10\hbar\omega_\perp$, we have also performed  
simulations with  $\mu=5\hbar\omega_\perp$. The corresponding dynamics are very similar and the sole 
discriminant parameter between the distinct vortex interaction 
regimes is indeed the orbit parameter $\chi$. The critical value 
$\chi_c$ switching from rebound to double reconnection dynamics is 
$0.25<\chi_c<0.28$ for $\mu=5$ and $0.22< \chi_c <0.25$ for $\mu=10$, 
supporting our argument that the value of $\mu$ does not change the essence of the physics.

\section{\label{interpretation}Interpretation of the results}

\subsection{Rebounds}

The simulations in Fig.~\ref{fig:4}a show that
rebound events are characterized by non intersecting vortex trajectories,
as we observe experimentally in a  
subset of images, e.g. in Fig.~\ref{fig:3}a,b. For example, 
Fig.~\ref{fig:4}a can be related to Fig.~\ref{fig:3}b, where the 
orientations extracted from the residuals start from an orthogonal 
configuration before partially overlapping
(however the trajectories do  not intersect)
and then emerge later showing an anti-parallel configuration.
A simpler, non-rotational, bounce is the one in Fig.~\ref{fig:3}a, where 
vortices are already anti-parallel before interacting.
Both of the observed rebounds 
are characterized by an increased visibility
when vortices are very close to each other. This is because
the residuals are generated by subtracting the
unperturbed density distributions and vortices become
more visible where their cores lie within a region of higher density.
The observed increase of  vortex visibility in rebound events is thus
consistent with the radial drift of the vortices 
towards the $x$-axis seen in numerical simulations.

\begin{figure*}[t]
\includegraphics[width=\textwidth]{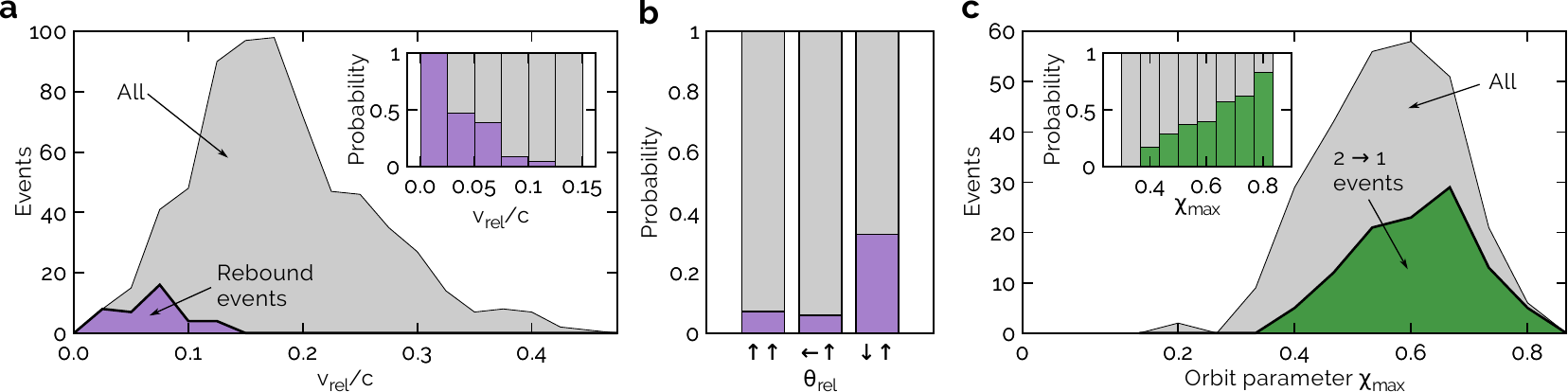}
\caption{Statistical analysis of experimental observations. 
(a) occurrence of rebound events (purple) as a function of 
the vortex-vortex relative velocity, within the ensemble of all 
collision events (grey). The velocity $v_{\mathrm{rel}}$ is normalized to the speed 
of sound $c$ evaluated at the center of the BEC. The inset shows the relative occurrence for 
each bin; (b) fraction of rebound events as a function of the 
relative angle just before the interaction; (c) occurrence of 
events (green) in which one vortex line disappears after the interaction, 
as a function of the largest orbit parameter of the vortex pair $\chi_{\text{max}}$, 
\textit{i.e.}, the amplitude of the outer vortex 
orbit in the BEC; the inset shows the relative occurrence per bin.}
\label{fig:5}
\end{figure*}

By studying the dynamics of hundreds of different experimental realisations, 
we make a statistical analysis which reinforces our interpretation. 
Figure \ref{fig:5}a shows the distribution of events as a function of the 
relative axial velocity of two approaching vortices. 
It is evident that those events, that are identified as rebounds (with approaching, but not touching, 
trajectories), preferentially happen when the relative velocity is small.  
As anticipated, the relative  angle $\theta_{\mathrm{rel}}$ in the radial plane matters when 
discerning rebound events from reconnections. Fig. \ref{fig:5}b shows 
the rebound probability as a function of the vortex relative angle just before their approach.
In order to classify the events in the three bins of Fig. \ref{fig:5}b,  we
use the relation between the shape of the residuals and the orientation of the vortex as extracted 
from numerical simulations (see Fig.~\ref{Fig:Delta_Ori} in Appendix C) to post-select all collisions 
for which we can safely estimate the relative angle to be approximately $0$, $45$, and $90$ degrees, 
within an uncertainty of the order of $\sim 30$ degrees. Then, in each group we count the fraction of 
rebounds.  The results confirm that rebounds are most likely to occur between anti-aligned vortices, consistent 
with the simulations.

\subsection{Orbiting dynamics} 
Two parallel vortices can orbit around the center of 
the BEC in the same direction with distinct orbit parameters $\chi$, 
only weakly interacting when they are at the closest distance.  
When imaged from a radial direction, the two vortices 
appear to cross periodically; in reality, they pass by each other
without visible changes of the residual pattern,
with, at most, only slight modifications of orbits and visibility. An example of
such orbiting dynamics can be observed in the 
experimental image Fig.~\ref{fig:3}c, and a similar case in the numerical 
simulations is shown in  Fig.~\ref{fig:4}e. 

\subsection{Reconnections} 
If the initial orientations of the vortices are not parallel and the axial collision dynamics is 
sufficiently fast, single reconnection processes are favored. As simulations show 
(Fig.~\ref{fig:4}c,d), these reconnection processes generate cusps which, as they relax, 
form Kelvin waves \cite{Kivotides2001}, {\it i.e.} helical perturbations of the cores, as for instance
the ones visible at $\tau=0.192$ in Fig.~\ref{fig:4}c. The excitation of Kelvin waves {\it via} vortex
reconnections was observed in superfluid Helium \cite{Fonda14} and similar effects have 
also been found in numerical simulations of Fermi superfluids \cite{Wlazlowski15}). 
In our experiment, such a perturbation of the vortex lines in a reconnection event implies 
a sudden change of both the orbit and the residual pattern, along with a 
significant change of visibility of one or both vortices, as illustrated in Fig.~\ref{fig:3}d,e.  
The nonlinear interaction among Kelvin waves might lead to Kelvin 
wave cascades \cite{Kozik04,Kozik06}. However, in the confined geometry of our elongated
BEC, the role of Kelvin waves is expected to be reduced compared to a uniform superfluid, 
due to finite (transverse) size effects. This is consistent with the fact that, if we release the 
whole condensate from the trap in order to observe the vortex lines by taking absorption
images in the axial direction, as done in \cite{Donadello14,Serafini15}, we  
typically observe almost rectilinear vortices with only smooth bends (mostly 
induced by boundary conditions for off-centered vortices), even in the presence of 
two or more vortices in the condensate.

\subsection{Ejections} 
When a vortex orbits the outer part of the condensate  (large $\chi$)  
a fast interaction with another vortex (either {\it via} 
a reconnection or a close orbiting interaction)
can cause the expulsion of either vortex towards the 
surface of the BEC, where the density is too low for observation
(the other vortex remaining inside). Examples of such ejection processes are 
shown in  Fig.~\ref{fig:3}e,f (reconnection-induced and 
orbiting-induced, respectively). The numerical counterpart of Fig.~\ref{fig:3}e is 
illustrated in Fig.~\ref{fig:4}d
(in order to emphasize this vortex-visibility effect in the numerical simulations, 
the thickness and the color of the lines in the plots reported in the right column of Fig.~\ref{fig:4} 
are modulated by the Thomas--Fermi density at which the corresponding vortex core resides, see Appendix B). 
A statistical analysis of experimental data is given in Fig.~\ref{fig:5}c:  
excluding cases where rebounds occur, we count all events of vortex-vortex interaction as a 
function of the largest orbit parameter of the vortex pair $\chi_{\text{max}}$. 
Then, among them, we show in green those in which the visibility of one of the two vortex lines is lost in the 
interaction. The relative distribution in the inset clearly supports the idea that ejections 
occur at large $\chi$, \textit{i.e.}, near the boundary of the condensate, in agreement with the 
result of the numerical simulations. These ejection processes might play a key role in the early post-quench 
dynamics of the BEC, when most of the vorticity produced by the Kibble--Zurek mechanism is 
progressively lost at the boundaries, eventually leaving only a few vortex lines in the final BEC \cite{Liu16}. 
 It is also worth noticing that a similar dynamics was previously discussed in Ref.~\cite{Becker13}. In 
 that case, pairs of dark solitons are created by an optical phase imprinting technique and their subsequent 
 dynamics  is observed. GP simulations show that solitons first decay into vortex rings and then into pairs of
 solitonic vortices which, in the experimental conditions, are still detected as dark soliton stripes. Hence
 a collision between two soliton stripes is actually a collision between two pairs of vortices.  Such 
 collisions can be inelastic and can also lead to ``sling shot" events where one of the solitonic vortices is 
 ejected from the condensate. Due to the different mechanism for the creation of vortices, the configurations
 discussed in Ref.~\cite{Becker13} involve typically more than two vortices in each collision, and thus 
 the dynamics is more complex than in our case, though qualitatively consistent. 

\section{\label{conclusions}Conclusions}
In conclusion, we have developed an innovative experimental
technique which, combined with numerical simulations,
is capable of determining the real-time
position and 3D orientation of vortex lines
in an elongated BEC. This combined technique
allows us to investigate vortex
dynamics in a 3D quantum system with unprecedented
resolution:
novel types of vortex interaction regimes are unambiguously identified beyond standard reconnections
already observed in superfluid helium \cite{Bewley08}.
While in uniform, unbounded and non-rotating superfluids reconnections of vortex lines 
moving towards each other are unavoidable,
and their effects have been extensively investigated
\cite{Koplik93,Nazarenko03,Dewaele94,Tebbs11,Zuccher12,Kerr11,Villois16b,Rorai16},
here we show that in a confined and inhomogeneous superfluid, 
depending on the relative velocity and orientation, 
two vortex lines can also rebound, perform double reconnections, 
maintain their orbits with negligible interaction and
undergo ejections. 
These  processes should play
even more important roles when the BEC contains more than two
vortices, for example in the case of turbulence \cite{Tsatsos16}.

\bigskip


{\bf Acknowledgments}
We acknowledge useful discussions with Fabrizio Larcher, Nick Proukakis, George Stagg and Marek Tylutki.
L.G.'s work is supported by Fonds National de la Recherche, Luxembourg, Grant n.7745104
and by the Italian National 
Group of Mathematical Physics (GNFM-INdAM).
C.F.B. acknowledges grant EPSRC EP/I019413/1.
T.B. and F.D. acknowledge the EU QUIC project for financial support.
The work was financially supported also by Provincia Autonoma di Trento.
S.S. and L.G. equally contributed to the work.\\


\section*{Appendix A: Experimental procedure for real-time imaging}

\begin{figure}[t!]
\includegraphics[width=\columnwidth]{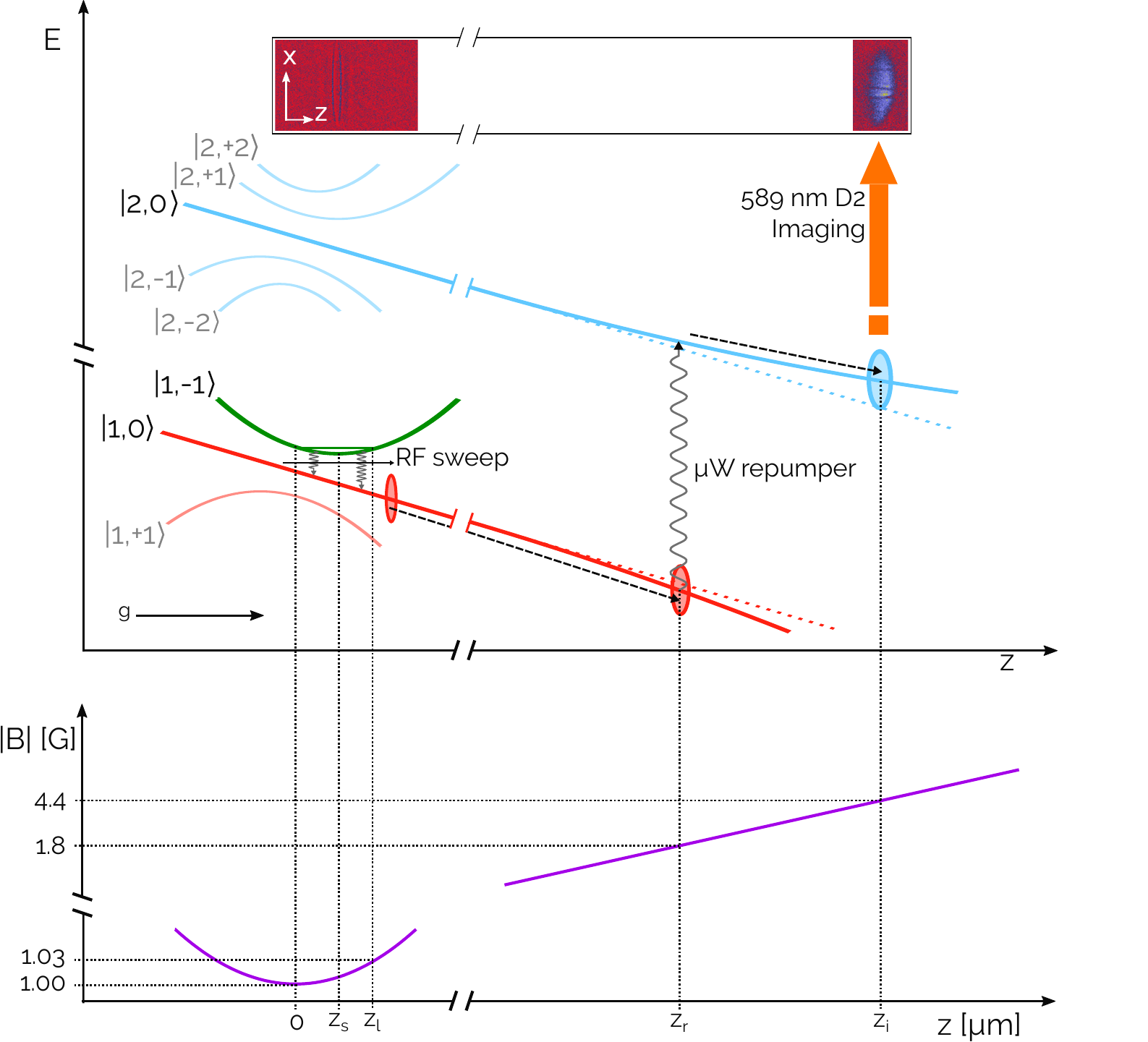}
\caption{Schematic picture of the outcoupling technique.  In the upper part of the figure, energy levels are reported as a function of the vertical coordinate. In the lower part, the modulus of the trapping magnetic field is reported. The sketched outcoupled atoms in red and cyan are not to scale.}
\label{fig:6}
\end{figure}

Atoms are trapped in a harmonic magnetic trap. The presence of gravity adds a linear potential that shifts the total potential minimum, $z_s$, 30 $\mu$m below the magnetic field minimum. This makes the BEC lie in a region of inhomogeneous magnetic field, mainly varying in the vertical direction $z$.
Figure \ref{fig:6} illustrates how the different hyperfine energy levels vary in space because of the second order Zeeman effect.

The \textit{rf} field couples the trapped state $|1,-1\rangle$ (green) with the non-magnetic state $|1,0\rangle$ (red). 
The resonant frequency is scanned linearly in time from above the BEC to below, at 10 kHz/ms.
The position $z_l=z_s+R_{\perp}\approx 47\,\mu$m corresponds  to the lowest boundary of the condensate.

The \textit{rf }sweep extracts each time a very small fraction of atoms, $\Delta N/N_0 \approx 1\%$ and the extraction process is then iterated many times in order to extract the vortex dynamics in real-time. 
The extracted atoms expand and fall freely under the effect of gravity. 
A microwave field is continuously kept on to couple $|1,0\rangle$ and $|2,0\rangle$ at the position $z_r\approx 280\,\mu$m, far enough from the trapped BEC. In this way the extracted falling atoms are transferred to $|2,0\rangle$ as soon as they cross such a surface and become detectable 
with the D2 probe light, as sketched in Fig.~\ref{fig:6}. 
We probe the extracted atoms via standard absorption imaging after $13$~ms of
total time of flight at $z_i\approx 830\,\mu$m from the trap center. 
An example of experimental image is shown in the inset on the top: only a weak diffraction pattern is visible in the trap region, while the outcoupled atoms become visible through absorption imaging below the repumper surface for $z>z_r$, when the sample is promoted to the bright state (cyan).

Experimental images are digitally filtered through a FFT analysis to remove fringes due to the optics elements along the imaging path. The two-dimensional optical density matrix of the sample is integrated along the vertical radial axis $z$ obtaining a one-dimensional axial profile. Such a profile is then fitted with a fourth-order polynomial and residuals are calculated. This procedure is performed on each extraction and then the full temporal sequence is reconstructed in order to follow the vortex trajectories in the trapped condensate. Each sequence is plotted using both a sequential and a diverging color map, to highlight respectively the trajectory of vortices and the pattern in the residuals resulting from the self-interference of outcoupled atoms. The axial position of the vortex gives us direct information on the amplitude of the orbit, and hence the orbit parameter $\chi$, as well as on the vortex velocity. \\

\section*{Appendix B: Numerical simulations of in-trap dynamics}

\begin{figure*}[ht!]
\includegraphics[width=1\textwidth]{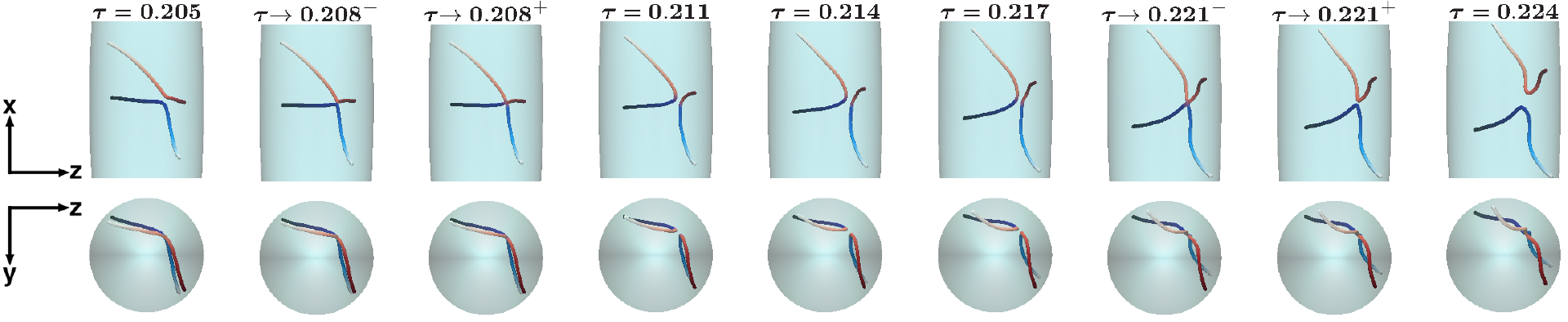}
\caption{Temporal sequence of radial (top) and axial (bottom) snapshots from GP simulations showing
in detail two interacting vortex lines undergoing a double reconnection. The snapshots refer to the numerical simulation reported in Fig. 4b, here illustrated employing a finer temporal resolution.
} 
\label{Fig:Double_Recon}
\end{figure*}

Real-time dynamical simulations of a harmonically trapped BEC at $T=0$ are
performed  by solving the mean-field Gross--Pitaevskii (GP) equation  
\begin{equation}
i \partial_t \Psi = \left[ - \frac{1}{2} \nabla^2 +  \frac{1}{2} ( r_\perp^2 + \lambda^{2} x^2 )
+  {\tilde g} | \Psi |^2  \right] \Psi
\nonumber
\end{equation}
for the complex macroscopic wave function of the condensate $\Psi=\Psi_\Re + i\Psi_\Im$.
Here  $r_\perp= (y^2 + z^2)^{1/2}$ is the radial 
coordinate, and $\lambda=\omega_x/\omega_\perp$ is 
the ratio of the axial to radial trapping frequencies. The mean-field coupling constant 
is ${\tilde g}=4\pi N a_s / \ell$,  where $a_s$ is the $s$-wave scattering length and  
$\ell=\sqrt{\hbar/(m\omega_\perp)}$  is the radial harmonic oscillator length. 
In the simulations reported in Fig.~\ref{fig:4} we use $\lambda=0.2$ and ${\tilde g}=7.4\times 10^3$, 
which corresponds to  $R_\perp / \xi =2\mu/(\hbar \omega_\perp)=20$, 
where $\mu$ is the chemical potential and $\xi=\hbar/\sqrt{2m\mu}$ is
the healing length. 
With respect to the experimental set up, in the numerical simulations
$\lambda$ is twice as large and
the chemical potential $\mu$ is approximately three times smaller. This 
choice is dictated by the computational resources available, but the key
characteristics of the vortex dynamics remain unchanged. 

We start the simulation with a Thomas--Fermi parabolic profile 
for the condensate density $|\Psi|^2$. In order to calculate the vortex-free 
ground state, we
evolve the GP equation in imaginary time until the relative decrease of 
energy $\Delta E/E$ 
between two consecutive time-steps 
is smaller than the threshold $\epsilon=10^{-5}$. 
Once this ground state is reached, we numerically imprint the 
two vortices. 

For the numerical simulations illustrated in Fig.~\ref{fig:4}a-c,
the vortices are initially imprinted in an orthogonal configuration,
intersecting the $x$-axis at the points $(x_0,0,0)$ 
and $(-x_0,0,0)$, the first vortex being oriented in 
the positive $z$ direction, the second vortex in the negative $y$ 
direction. The corresponding values of the orbit parameter $\chi$
are $0.22$, $0.25$, $0.375$, for simulations reported in Fig.~\ref{fig:4}a,b,c respectively.
In the simulation illustrated in Fig.~\ref{fig:4}d, vortices are also initially
orthogonal, but the lower vortex is imprinted in the 
center of the BEC and oriented in the positive $y$ direction. The 
orbit parameter $\chi$ of the off-centered vortex is 0.7.
Concerning the last simulation, Fig.~\ref{fig:4}e, both vortices are 
oriented in the positive $y$ direction with $\chi=0.33,0.5$. 

Vortex imprinting
is accomplished by imposing a Pad\'{e} density 
profile \cite{Berloff04} and a $2\pi$ phase winding around the vortex axis.
We then let the system evolve in imaginary time towards the lowest energy 
state employing the previously described energy
convergence criterion. Once $\Delta E/E < \epsilon$, we start the 
evolution of the GPE in real time.

Our numerical code employs second-order accurate finite difference 
schemes to discretize spatial derivatives; the integration in time 
is performed via a 4-th order Runge--Kutta method. 
The grid-spacings are homogeneous in the 
three Cartesian directions  
($\Delta x=\Delta y=\Delta z=\xi/3=0.075\ \ell$) and the time step 
is $\Delta t = 0.00125\ \omega_\perp^{-1}$. The number of grid-points
in the $x$, $y$ and $z$ direction are  \{$N_x,N_y,N_z$\}=\{800,224,224\}, leading to a computational box 
$\left \{ \left [ x_{\rm min} : x_{\rm max} \right ] \times \left [ y_{\rm min} : y_{\rm max} \right ] \times \left [ z_{\rm min} : z_{\rm max} \right ] \right \} = 
\left \{ \left [ -30 : 30 \right ] \times  \left [ -8.4 : 8.4 \right ] \times \left [ -8.4 : 8.4 \right ] \right \}$, where these values are expressed in units of $\ell$.

Vortex tracking is achieved via an algorithm based on the 
pseudo-vorticity unit vector 
\begin{equation}
\hat{\bm{\omega}}~:=~\frac{\nabla \Psi_\Re \times \nabla \Psi_\Im}{\left | \nabla \Psi_\Re \times \nabla \Psi_\Im \right |}
\nonumber
\end{equation}
which is tangent to the vortex line along its length~\cite{Rorai16,Villois16}. 
To identify the first (starting) point along the axis of each vortex, 
we use criteria based simultaneously on circulation and density, and then 
adopt a steepest descent algorithm to achieve sub-grid resolution.
Successive points on vortex lines are determined with separation distance 
$\Delta \zeta=\Delta z /10$.

In the plots reported in the right column of Fig.~4, the initial line colors refer to the colors 
of the vortices illustrated in the snapshots (red/blue) until a reconnection event occurs.
After the latter, the colors employed switch to orange/green and again to red/blue
if a second reconnection takes place. 
The transparency of the lines and the intensity of the colors employed in the plots
aim to reproduce the expected experimental vortex visibility. To achieve this aim we proceed
as follows. Since the experimental visibility is obtained by subtracting the background (vortex-free) 
density profile from the optical integrated densities, the vortex visibility
increases for increasing atom number depletion arising from the presence of the vortex itself. 
As a consequence, both width and color of the lines plotted in the right column of Fig.~4
are weighted by the value of the Thomas-Fermi density evaluated at the centre of vorticity 
of the corresponding vortex,
in order to account, at least qualitatively, for the actual visibility of the vortex in the residual. \\

\noindent
Finally, in order to illustrate  in more detail the
double reconnection dynamics, in Fig.~\ref{Fig:Double_Recon} we report 
radial and axial snapshots of vortex configurations for the double
reconnection event already described in Fig.~\ref{fig:4}b, but employing a finer temporal resolution. The exchange of vortex strands and
the formation of cusps is clearly visible at the reconnection instants $\tau = 0.208$ and $\tau = 0.221$. After the second
reconnection evidence for the formation of Kelvin waves can be observed (at $\tau = 0.224$).

\section*{Appendix C: Numerical simulations of expansion}

\begin{figure}[b!]
\includegraphics[width=1\columnwidth]{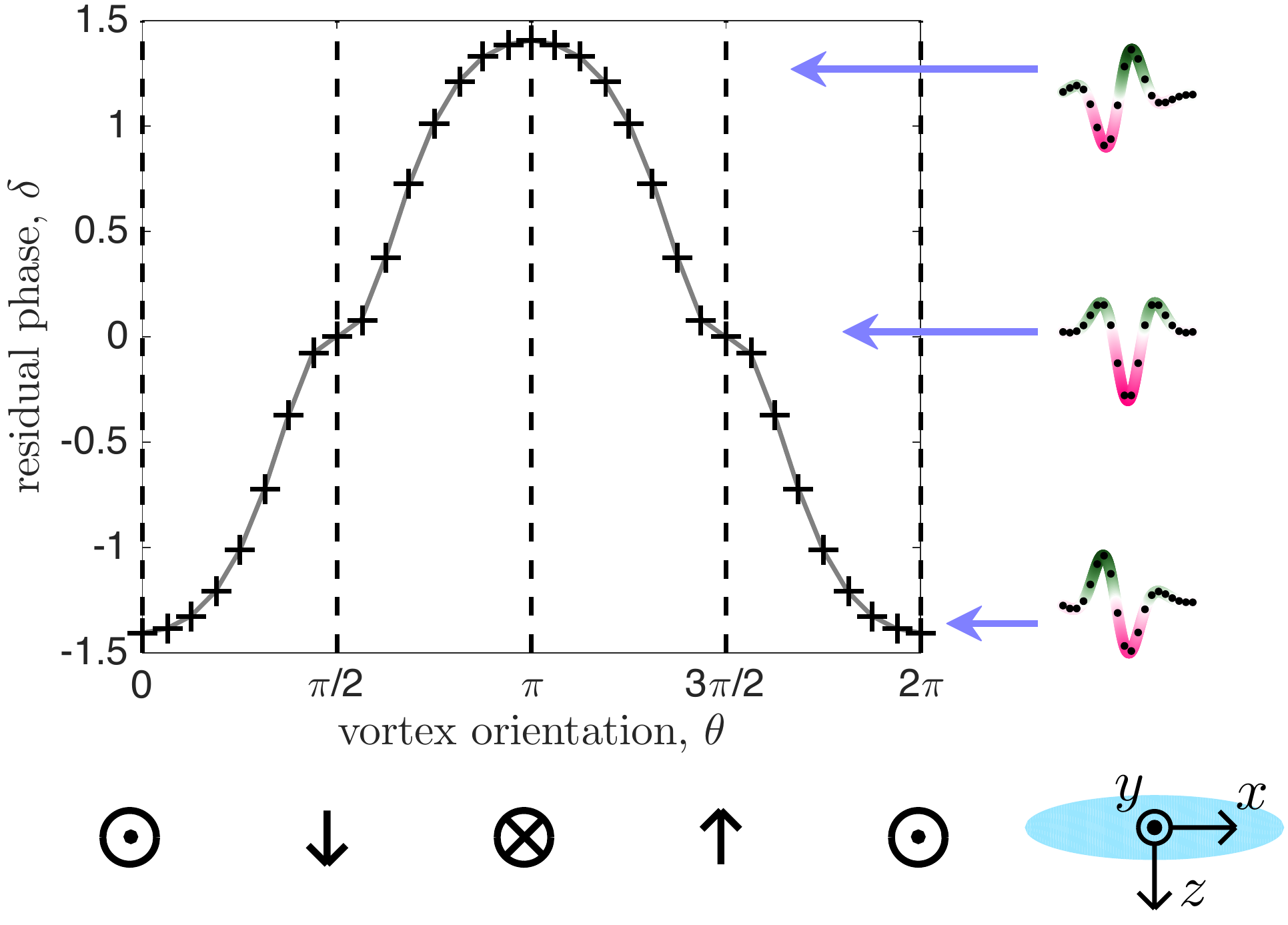}
\caption{Residual phase $\delta$ of the extracted portion, after a 13 ms expansion, for various {\it in situ} vortex orientations $\theta$, obtained by using the function (\ref{eq:fit}) to fit the shape of the residuals in GP simulations. These numerical results are for a solitonic vortex stationary state that passes through the condensate center, i.e.~even though the vortex is considered for numerous orientations it is always a straight line that remains in the $x=0$ plane. The (+) symbols are the data points while the solid line guides the eye. For reference, the vertical dashed lines indicate examples of the {\it in situ} vortex orientation as indicated by the lower arrows.
} 
\label{Fig:Delta_Ori}
\end{figure}

For the purpose of inferring \emph{in situ} vortex information from the post-expansion 
residual densities, we explicitly simulate the outcoupling and expansion dynamically 
using the GP equation.
The outcoupled atoms expand as they fall under gravity relative to the trapped condensate.
While the outcoupled atoms may be fairly dilute, they still experience significant 
interactions with the dense trapped portion, for several ms, until gravity finally 
separates the components.
The corresponding scattering length between outcoupled and trapped atoms is the same as that between trapped atoms, and takes the value 54.54(20)$a_0$, where $a_0$ is the Bohr radius. Interactions between outcoupled atoms, while less important, are also included and for these the scattering length is 52.66(40)$a_0$ \cite{Knoop11}.

The partial extraction is performed as a linear-in-energy sweep such that upper atoms are outcoupled before lower atoms.
The phase of the trapped component is allowed to evolve during the sweep, 
which occurs over a few ms, but the in-trap vortex dynamics is much slower and we treat this as fixed.
The trapped component's phase, owing to a larger potential energy, evolves more rapidly during the sweep than it does for the released atoms.
Consequently, if we consider the example of a horizontal vortex ( {\it i.e.} oriented in the $y$ direction), by the time the lower atoms are released they have accumulated a greater phase change than the upper ones, which were released earlier, such that these layers interfere constructively on one side of the core and destructively on the other, depending on the sense of the {\it in situ} phase circulation.
Furthermore, the combined effects of gravity and the intercomponent interactions mean that the speed  of the sweep is important.
We choose a sweep speed, both experimentally and theoretically, which rapidly compresses 
the outcoupled cloud in the vertical direction, thus maximizing interference effects. This enhances the $x$-direction asymmetry of the residual, allowing us to determine the orientation and, for horizontal alignment,  the sign of the  vortex.
When extracting vortex information and quantifying this asymmetry we fit the function
\begin{equation}
f_{\rm fit}(x) = \frac{A\cos[B(x-x_v)+\delta]}{\cosh^2\left[(x-x_v)/C\right]}
\label{eq:fit}
\end{equation}
to the 1D residual,  where $A, B, C, x_v$ and $\delta$ are fitting parameters. Here, $x_v$ 
represents the axial position of the vortex while $\delta$ is related to the orientation  
$\theta$ of the vortex line in the radial plane.

To ensure numerical convergence of the residual to $\sim$1\% when performing a $13$~ms 
expansion we begin with a grid that initially represents the \emph{in situ} density, \{$x_{\rm max}$,$y_{\rm max}$,$z_{\rm max}$\} = \{51,16,14\} $\ell$ and \{$N_x,N_y,N_z$\} = \{300,180,120\}, and after several interpolations, end with a much-enlarged grid, \{$x_{\rm max}$,$y_{\rm max}$,$z_{\rm max}$\} = \{75,70,60\} $\ell$ and \{$N_x,N_y,N_z$\} = \{180,180,600\}.

The relation $\delta(\theta)$ is numerically calculated for a straight solitonic vortex and the results are displayed in Fig.~\ref{Fig:Delta_Ori}. Importantly, the fitting function gives a negative value of $\delta$ for a horizontal vortex aligned in the $+y$ direction ($\theta=0$), whereas the sign of $\delta$ flips for a horizontal vortex of the opposite sense, i.e.~$\delta(\theta=\pi)=-\delta(\theta=0)$. For vertically oriented vortices ($\theta= \{\pi/2 , 3\pi/2\}$) one finds $\delta=0$, and we reiterate here that since every $\delta$ corresponds to two angles this method cannot, for example, determine the sense of a vertical vortex.
We note further that the relationship given by Fig.~\ref{Fig:Delta_Ori} is expected to be modified for vortices with large orbit parameters.


%

\end{document}